\documentclass[reqno,12pt]{article}
\usepackage{amsfonts,amssymb,
amsmath,amscd, bm, mathrsfs, 
mathrsfs,delarray,subfigure}
\usepackage{hyperref,cite}
\usepackage[all]{xy}
\usepackage[dvips]{color}
\usepackage[dvips]{graphicx}
\usepackage{pstricks}
\usepackage{pstricks-add}
\usepackage{pst-solides3d}

\DeclareMathOperator{\vol}{vol}
\DeclareMathOperator{\im}{im}
 
\DeclareMathOperator{\id}{id}

\DeclareMathOperator{\Map}{Map}

\DeclareMathOperator{\Gau}{Gau}

\DeclareMathOperator{\Aut}{Aut}

\DeclareMathOperator{\Inn}{Inn}

\DeclareMathOperator{\Diff}{Diff}

\DeclareMathOperator{\tr}{tr}

\DeclareMathOperator{\GL}{GL}

\DeclareMathOperator{\UU}{U}

\DeclareMathOperator{\MCG}{MCG}
\DeclareMathOperator{\Conj}{Conj}
\DeclareMathOperator{\Triv}{Triv}
\DeclareMathOperator{\Cyc}{Cyc}
\DeclareMathOperator{\Orb}{Orb}

\numberwithin{equation}{subsection} 
\numberwithin{subsection}{section} 

\newcommand{\varrhd}{{\,\vartriangleright\,}}
\newcommand{\varlhd}{{\,\vartriangleleft\,}}

\newcommand{\varsucc}{{\,\succ\,}}

\font\sansserif=cmss12
\font\scriptsansserif=cmss12 at 7 truept
\font\scriptscriptsansserif=cmss10 at 5 truept
\def\sans{\fam=14}
\newcommand{\mathsans}[1]{{{\sans #1}}}

\font\euler=eusm10 at 12.8 truept
\font\scripteuler=eusm7
\font\scriptscripteuler=eusm5 
\CompileMatrices

\newtheorem{defi}{
Definition}[subsection]

\newtheorem{prop}{
Proposition}[subsection]

\begin{document}

\hrule\vskip.5cm
\hbox to 14.5 truecm{May 2015 \hfil DIFA 15}
\vskip.5cm\hrule
\vskip.7cm
\centerline{\textcolor{blue}{\bf ON HIGHER HOLONOMY INVARIANTS}}   
\centerline{\textcolor{blue}{\bf IN HIGHER GAUGE THEORY II}}   
\vskip.2cm
\centerline{by}
\vskip.2cm
\centerline{\bf Roberto Zucchini}
\centerline{\it Dipartimento di Fisica ed Astronomia, Universit\`a di Bologna}
\centerline{\it V. Irnerio 46, I-40126 Bologna, Italy}
\centerline{\it I.N.F.N., sezione di Bologna, Italy}
\centerline{\it E--mail: emanuele.soncini@studio.unibo.it, zucchinir@bo.infn.it}
\vskip.7cm
\hrule
\vskip.6cm
\centerline{\bf Abstract} 
\par\noindent
 This is the second of a series of two technical papers devoted to the analysis of holonomy 
invariants in strict higher gauge theory with end applications in higher 
Chern--Simons theory. We provide a definition of trace over a crossed module
such to yield surface knot invariants upon application to $2$--holonomies. 
We show further that the properties of the trace are best described 
using the theory quandle crossed modules. 

 \vfil\eject

\tableofcontents

\vfil\eject


\vfil

\section{\normalsize \textcolor{blue}{Introduction}}\label{sec:intro}

\hspace{.5cm} 
The topological classification of ordinary knots in $3$--dimensions through their invariants is one 
of the outstanding problems of contemporary low dimensional topology \cite{Kauffman:1991ds,Kauffman:1995hc}.
Such issue has higher dimensional analogs, in particular the characterization of the topology of knotted surfaces 
in $4$--dimensions by means of suitable higher invariants \cite{Carter:1998cas,Kamada:2002ska}.

In the first paper of a series of two,
we pointed out that the success of $3$--dimensional Chern--Simons theory as a quantum field theoretic 
framework for the computation of ordinary knot invariants \cite{Witten:1988hf,Witten:1992fb}
suggests that a $4$--dimensional 
version of Chern--Simons theory may do the same with regard to surface knot invariants.
Since the higher dimensional analogs of plane Chern--Simons theory exist only in 
odd dimensional spaces, the realization of Chern--Simons theory appropriate for surface knots 
is likely to belong to the domain of higher gauge theory \cite{Kotov:2007nr,Fiorenza:2011jr}.
(See refs. \cite{Baez:2002jn,Baez:2010ya} for background information on the subject
and refs. \cite{Schreiber2011,Gruetzmann:2014ica} for a comprehensive in depth treatment.) 
In refs. \cite{Zucchini:2011aa,Soncini:2014ara}, a higher gauge theoretic $4$--dimensional 
Chern--Simons model was constructed resembling in many ways the usual Chern--Simons one
having the eventual calculation of surface knot invariants as its goal. 
An alternative approach relying on $BF$ theory instead has been pursued in \cite{Cattaneo:2002tk}.

In ordinary Chern--Simons theory, the computation of knot invariants 
involves the evaluation of traces of Wilson loops of the gauge field, mathematically holonomies 
of the gauge connection, along knots in representations of the gauge group \cite{Marino:2005sj}. 
In a strict higher gauge theory, the corresponding issue for surface knots
has two parts: $(a)$ the definition of surface holonomies and the analysis 
of their dependence on the choice of gauge and base data; $(b)$
the definition of the appropriate notion of trace for the gauge crossed module
yielding surface knot invariants upon application to surface holonomies. 
The first part has been treated in the first paper of the series \cite{SZ:2015}, the second one 
is the topic of the present second paper.  


\subsection{\normalsize \textcolor{blue}{Scope and plan of the paper}}\label{sec:scope}

\hspace{.5cm} 
We now illustrate the scope of our analysis in an illustrative way with no pretence to
mathematical rigour. 

In an ordinary gauge theory with gauge group $G$ and a trivial principal $G$--bundle
as background, one can associate with any flat connection
$\theta$ and based knot $\xi$ the holonomy $F_\theta(\xi)\in G$ \cite{SZ:2015}.
$F_\theta(\xi)$ is not gauge invariant and depends also on the location of the base point of $\xi$
in general. 
Smoothly changing the choice of the gauge and base data however affects $F_\theta(\xi)$ at most 
by a simple conjugation by some group element $a\in G$, viz
\begin{equation}
F_\theta(\xi) \rightarrow aF_\theta(\xi) a^{-1}.
\label{}
\end{equation}
Thus, for given $\theta$ and $\xi$, only the conjugation class of $F_\theta(\xi)$
is uniquely determined in a gauge and base independent manner. 
If we are to extract a numerical invariant out the holonomy of a knot, we need a 
trace over $G$, a class function in common parlance, 
that is a conjugation invariant mapping $\tr:G\rightarrow \mathbb{C}$, viz
\begin{equation}
\tr(aza^{-1})=\tr(z)
\label{}
\end{equation}
for $a,z\in G$. There is a well known procedure of construction of such functions. Given any representation 
$R:G\rightarrow \GL(X)$ in some complex vector space $X$, the mapping 
$\tr_R:G\rightarrow \mathbb{C}$ defined by $\tr_R(a)=\tr_X(R(a))$
for $a\in G$ has the above property.

In this paper, we study surface knots using strict higher gauge theory. 
A based surface knot $\varXi$ of genus $\ell_\varXi$ is characterized by an assignment 
of a point and $2\ell_\varXi$ independent non contractible loops $\zeta_{Mi}$ 
based at that point, called characteristics line knots of $\varXi$. 
In a strict higher gauge theory with gauge crossed module $(G,H)$ and a trivial principal 
$(G,H)$--$2$--bundle as background, one can associate with any flat $2$--connection
doublet $(\theta,\varUpsilon)$ and based surface knot $\varXi$ 
the surface holonomy $F_{\theta,\varUpsilon}(\varXi)\in H$ and 
the $2\ell_\varXi$ holonomies $F_{\theta,\varUpsilon}(\zeta_{Mi})\in G$ 
\cite{SZ:2015}. $F_{\theta,\varUpsilon}(\varXi)$ and the $F_{\theta,\varUpsilon}(\zeta_{Mi})$ 
are not $1$--gauge invariant and depend also on the location of the 
characteristic line knots of $\varXi$. Changing smoothly the choice of the gauge and base data however 
affects $F_{\theta,\varUpsilon}(\varXi)$, $F_{\theta,\varUpsilon}(\zeta_{Mi})$
by a joined $2$--conjugations by some crossed module elements $a\in G$ and $A_i\in H$, viz 
\begin{align}
&F_{\theta,\varUpsilon}(\varXi)\rightarrow m(a)(F_{\theta,\varUpsilon}(\varXi)),
\vphantom{\Big]}
\label{}
\\
&F_{\theta,\varUpsilon}(\zeta_{Mi})\rightarrow aF_{\theta,\varUpsilon}(\zeta_{Mi})a^{-1}t(A_i),
\vphantom{\Big]}
\label{}
\end{align}
where $t$ and $m$ are target and action maps of the crossed module $(G,H)$. 
Therefore, for given $(\theta,\varUpsilon)$ and $\varXi$, only the joined $2$--conjugation class of 
$F_{\theta,\varUpsilon}(\varXi)$ and the $F_{\theta,\varUpsilon}(\zeta_{Mi})$ is 
uniquely determined in a $1$--gauge and base independent manner. 
If we are to distill numerical invariants out the holonomies of a surface knot, we need a 
$2$--trace over $(G,H)$, which we define as a pair of mappings $\tr_b:G\rightarrow \mathbb{C}$, 
$\tr_f:H\rightarrow \mathbb{C}$ invariant under  $2$--conjugation, viz
\begin{align}
&\tr_b(aza^{-1}t(A))=\tr_b(z),
\vphantom{\Big]}
\label{}
\\
&\tr_f(m(a)(Z))=\tr_f(Z)
\vphantom{\Big]}
\label{}
\end{align}
for $a,z\in G$ and $A,Z\in H$. Unlike the ordinary case,
there is no standard procedure of construction of such functions
to the best of our knowledge. In this paper, we propose one that parallels as much as possible 
the familiar one of ordinary gauge theory. 

In the above account, we have introduced at a somewhat elementary level
the notion of trace over a group $G$ or $2$--trace over a crossed module $(G,H)$. 
The problem of constructing such traces can be dealt with formally
using quandle theory. 

Quandles \pagebreak are algebraic structures which emerged in knot theory
\cite{Joyce:1982qa,Matveev:1984qt} in the 1980's. 
With every knot, there is associated a fundamental quandle. This is a complete knot invariant 
in the sense that two non oriented knots are topologically equivalent if and only if their
fundamental quandles are isomorphic. 
Ascertaining whether two given quandles are isomorphic is however invariably a 
prohibitively difficult task, so that the usefulness of the fundamental quandle as invariant
is limited. (See refs. \cite{Carter:2009qi,Carter:2012qc} for a readable introduction to the subject.) 

In this paper, we are going to use exploit quandle theory in a completely different way. 
What does really matter in the construction of holonomy invariants is the conjugation structure of 
the gauge group $G$, in ordinary gauge theory, 
and the $2$--conjugation structure of gauge crossed module $(G,H)$, in higher gauge theory. 
Those structures are captured by the 
conjugation quandle $\Conj(G)$ of $G$ in the former case and the conjugation
quandle crossed module $\Conj(G,H)$ of $(G,H)$ in the second one. 
Just as a trace over $G$ can be viewed as a morphism of the quandle $\Conj(G)$ 
into the trivial quandle $\mathbb{C}$ of the complex numbers, a $2$--trace over a $(G,H)$ 
can be viewed as a morphism of quandle crossed module $\Conj(G,H)$ into the 
trivial quandle crossed module $(\mathbb{C}_b, \mathbb{C}_f)$ of the pairs of complex numbers. 
This abstract approach may be a little overdone for complex valued traces, but it has the advantage 
of being immediately generalizable to more general target quandle structures.

A categorification of the notion of trace of endomorphisms of vector spaces exists.
(See refs. \cite{Selinger:2009ct,Ponto:2011ct} for a survey.)
In the most general version, it applies to braided monoidal categories
such as the category $\Bbbk$--Vect of vector spaces on a field $\Bbbk$. 
However, being inspired by the familiar linear algebraic concept, it is biased 
by it at its heart and so does not really serve our purposes. What does really matter here is 
the conjugation structure of groups and crossed modules, as we pointed out above, and this is captured 
by the quandle formulation. 

Just as the construction of invariant traces in a gauge theory with gauge group $G$ involves the choice of 
a representation of $G$, the construction of invariant $2$--traces in a higher gauge theory with gauge 
crossed module $(G,H)$ requires a representation of $(G,H)$. A representation of $(G,H)$
is traditionally defined as a strict $2$--functor from the delooping $BV$ of the strict 
$2$--group $V$ equivalent $(G,H)$, a $2$--groupoid, into the $2$--category $2$--$\Bbbk$--Vect 
of $2$--vector spaces on a field $\Bbbk$, whichever way they are defined. 
This allows to analyze all representations of $V$ on the same footing. We have found more natural to define
a representation of $(G,H)$ as a crossed module morphism of $(G,H)$ into a crossed module 
$(G',H')$ such that $G'$, $H'$ are subgroups of the general linear groups $\GL(X)$, $\GL(Y)$ 
for vector spaces $X$, $Y$. 


The basic definitions and results of quandle theory which we are going to use are 
expounded in sect. \ref{sec:quandle}. Our construction of $2$--traces is detailed in sect. 
\ref{sec:holoinv}.

\subsection{\normalsize \textcolor{blue}{Outlook}}\label{sec:outlook}

\hspace{.5cm} The symmetry of the higher $4$--dimensional Chern--Simons model of refs. 
\cite{Zucchini:2011aa,Soncini:2014ara} is based on a finite dimensional 
semistrict $2$--term $L_\infty$ algebra 
$\mathfrak{v}$ obeying certain conditions. In general, $\mathfrak{v}$ does not integrate 
to some kind of finite dimensional $2$--group, let alone a strict $2$--group 
equivalent to some crossed module. Much of the symmetry properties of the model,
however, can be phrased in terms of the automorphism group $\Aut(\mathfrak{v})$ 
of $\mathfrak{v}$, which is a finite dimensional strict $2$--group 
even though $\mathfrak{v}$ is merely semistrict and is therefore equivalent to a crossed module. 
It is conceivable, therefore, 
that holonomy invariants of surface knots may be computed in this model,
at least in principle, using the results of the present work. 

\vspace{1cm}

\noindent
\textcolor{blue}{Acknowledgements.}
We thank E. Soncini for participating in the early stages of this project and P. Ritter for useful discussions.
We acknowledge financial support from INFN Research Agency.

\vfil\eject

\section{\normalsize \textcolor{blue}{Quandle theory}}\label{sec:quandle}

\hspace{.5cm} Quandle theory is a well developed subject of abstract algebra
with a wide range of applications, especially in knot theory but also to other fields
of mathematics. In this section, we illustrate certain results of quandle theory, which 
will be applied in next section in our treatment of holonomy invariants. 
In subsects. \ref{sec:qu}, \ref{sec:qumor}, we recall the basic facts of quandle theory.
Good reviews on this topic are found in \cite{Carter:2009qi,Kamada:2002qr,Crans:2002lq}.
In subsects. \ref{sec:qucros}, \ref{sec:qucromor}, we study in some detail 
quandle crossed modules and their morphisms. The idea of crossed modules of quandles 
is taken from \cite{Crans:2013cmr} while the notion of augmentation which we 
introduce is originally ours to the best of our knowledge.


\subsection{\normalsize \textcolor{blue}{Quandles}}\label{sec:qu}

\hspace{.5cm} Quandles are algebraic structures abstracting the notion of conjugation. 

\begin{defi} \label{def:qu1}
A quandle is a set $Q\not=\emptyset$ with a binary operation $\varrhd:Q\times Q$ $\rightarrow Q$ 
with the following properties,
\begin{subequations}
\begin{align}
&a\varrhd a=a,
\vphantom{\Big]}
\label{qu1}
\\
&a\varrhd(b\varrhd c)=(a\varrhd b)\varrhd(a\varrhd c)
\vphantom{\Big]}
\label{qu2}
\end{align}
\end{subequations}
for any $a,b,c\in Q$. Moreover, for any $a,b\in Q$, the equation $a\varrhd c=b$ has a unique solution $c\in Q$
\end{defi}
The solution $c$ of $a\varrhd c=b$ is denoted as $c=a\varlhd b$. It can be shown that $(Q,\varlhd)$ is also a 
quandle. 

\begin{defi} \label{def:qu2}
A quandle $Q$ is said pointed, if there exists a distinguished element 
$1_Q\in Q$ such that \hphantom{xxxxxxxxxxxxxxxx}
\begin{subequations}
\begin{align}
&a\varrhd 1_Q=1_Q,
\vphantom{\Big]}
\label{qu3}
\\
&1_Q\varrhd a=a
\vphantom{\Big]}
\label{qu4}
\end{align}
\end{subequations}
for arbitrary $a\in Q$. $1_Q$ is called the neutral element of $Q$. 
\end{defi}

The prototypical pointed quandle is the quandle of a group.

\begin{prop} \label{prop:qu1}
Let $Q$ be a group. Define a mapping $\varrhd:Q\times Q\rightarrow Q$ by
\begin{equation}
a\varrhd b=aba^{-1},
\label{qu5}
\end{equation} 
where $a,b\in Q$. Then, $Q$ with its neutral element $1_Q$ is a pointed quandle. 
\end{prop}

\noindent {\it Proof}. From \eqref{qu5}, the axioms 
\eqref{qu1}--\eqref{qu4} are easily verified. \hfill $\Box$

\noindent
This quandle is called the conjugation quandle of the group $Q$ 
and is denoted by $\Conj(Q)$, when one wants to 
emphasize its being a quandle rather than a group. 

A set $Q\not=\emptyset$ with a distinguished element $1_Q$ is said pointed. 
With a pointed set, there is associated a basic quandle. 

\begin{prop} \label{prop:qu2}
Let $Q\not=\emptyset$ be a pointed set. Define a mapping $\varrhd:Q\times Q\rightarrow Q$ by
the relation \hphantom{xxxxxxxxxxxxxxxxxxx}
\begin{equation}
a\varrhd b=b,
\label{qu6}
\end{equation} 
where $a,b\in Q$. Then, $Q$ with its distinguished element $1_Q$ is a pointed quandle.
\end{prop}

\noindent {\it Proof}. By \eqref{qu6}, the axioms 
\eqref{qu1}--\eqref{qu4} hold trivially. \hfill $\Box$

\noindent
This quandle is called the trivial quandle of the set $Q$
and is denoted by $\Triv(Q)$, when one wishes to 
emphasize its having a quandle structure. 
It derives its name from the property that the mapping
$a\varrhd\cdot:Q\rightarrow Q$ is the identity $\id_Q$ for any $a\in Q$.  
The conjugation quandle of an Abelian group $Q$ is clearly trivial.


\subsection{\normalsize \textcolor{blue}{Morphisms of quandles}}\label{sec:qumor}

\hspace{.5cm} Quandle morphisms are mappings between quandles compatible with 
their quandle structures. 

\begin{defi} \label{def:qumor1}
Let $Q,Q'$ be quandles and let $\phi:Q\rightarrow Q'$ be a mapping. $\phi$ is a quandle
morphism if \hphantom{xxxxxxxxxxxx}
\begin{equation}
\phi(a\varrhd b)=\phi(a)\varrhd\!{}'\,\phi(b)
\label{qumor1}
\end{equation}
for arbitrary $a,b\in Q$. If $Q$, $Q'$ are both pointed, it is further required that
\begin{equation}
\phi(1_Q)=1_{Q'}. 
\label{qumor2}
\end{equation}
\end{defi}


\noindent
Quandles then organize as a category. 

\begin{prop} \label{prop:qumor1}
Quandles and quandle morphisms form a category. Pointed quandle and quandle 
morphism constitute a subcategory of it. 
\end{prop}

\noindent
{\it Proof}. From \eqref{qumor1}, it is easily verified that the composition $\phi'\circ \phi:
Q\rightarrow Q''$ of two quandle morphisms $\phi:Q\rightarrow Q'$, $\phi':Q'\rightarrow Q''$
the identity $\id_Q$ of a quandle are quandle morphisms. Therefore, quandles and quandle morphisms
constitute a subcategory of the category Set, hence a category itself. 
Similar conclusions are reached in the pointed case.
\hfill $\Box$

We shall denote by $\mathsans{Q}$ (resp. $\mathsans{PtQ}$) the category of 
ordinary (resp. pointed) quandles and quandle morphisms. 

\begin{prop}
Let $Q$ be a (pointed) quandle. For  $a\in Q$, let $\iota_a:Q\rightarrow Q$ be 
the mapping defined by the expression \hphantom{xxxxxxxxxxxxxx}
\begin{equation}
\iota_a(b)= a\varrhd b
\label{qumor3}
\end{equation}
with $b\in Q$. Then, $\iota_a$ is a 
(pointed) quandle automorphism.
\end{prop}

\noindent
{\it Proof}. First of all, we note that the map $\iota_a$ is invertible by one of the defining 
properties of a quandle structure (cf. def. \ref{def:qu1}). 
By \eqref{qu2}, further, we have 
$\iota_a(b\varrhd c)=a\varrhd (b\varrhd c)=(a\varrhd b)\varrhd(a\varrhd c)
=\iota_a(b)\varrhd\iota_a(c)$, showing that $\iota_a$ is a automorphism of the
quandles $Q$, as claimed.  \hfill $\Box$

\begin{defi}
Let $Q$ be a (pointed) quandle. 
An quandle automorphism of $Q$ is called inner if it is of the form
$\iota_a$ for some $a\in Q$.  
\end{defi}

Quandle inner automorphisms clearly answer to familiar group automorphisms. 
We let $\Inn(Q)$ be the subgroup of the (pointed) automorphism group 
$\Aut(Q)$ of $Q$ generated by its inner elements and their inverses.

\begin{prop}
Let $Q$ be a (pointed) quandle. Then, $\Inn(Q)$ is a normal subgroup of $\Aut(Q)$.  
\end{prop}

\noindent
{\it Proof}. Let $a\in Q$ and $\phi$ be an arbitrary quandle automorphism of $Q$. 
Using \eqref{qumor1}, we find 
that $\phi\circ\iota_a\circ \phi^{-1}(b)=\phi(a\varrhd \phi^{-1}(b))
=\phi(a)\varrhd b=\iota_{\phi(a)}(b)$
for $b\in Q$. As the subgroup $\Inn(Q)$ of $\Aut(Q)$ is generated by the automorphisms 
$\iota_a$ and their inverses, $\Inn(Q)$ is invariant, so normal in $\Aut(Q)$. \hfill $\Box$

Morphisms of groups induce morphisms of the associated pointed quandles
(cf. prop. \ref{prop:qu1}). 

\begin{prop} \label{prop:qumor2}
Let $Q,Q'$ be groups and $\phi:Q\rightarrow Q'$ be a group morphism. 
Then, $\phi$ is a morphism of the pointed quandles $Q$, $Q'$.  
\end{prop}

\noindent
{\it Proof}. From \eqref{qu5}, we have $\phi(a\varrhd b)=\phi(aba^{-1})
=\phi(a)\phi(b)\phi(a)^{-1}=\phi(a)\varrhd'$ $\phi(b)$
for $a, b$ $\in Q$ so that $\phi$ fulfills \eqref{qumor1}. 
The validity of \eqref{qumor2} is obvious, as $\phi$ is a group morphism.  \hfill $\Box$

If we denote $Q$, $Q'$ as $\Conj(Q)$, $\Conj(Q')$ to emphasize their being endowed with a quandle
structure, then the morphism $\phi$ gets denoted as $\Conj(\phi)$. The following proposition holds.

\begin{prop}
$\Conj$ is a functor from the category $\mathsans{Grp}$ of groups and group morphisms 
to the category $\mathsans{PtQ}$ of pointed quandles and pointed quandle morphisms. 
\end{prop}

\noindent
{\it Proof}. Indeed, compositions and identities are preserved by $\Conj$, as it is
 immediately checked.
\hfill $\Box$

A morphism $\phi:Q\rightarrow Q'$ of pointed sets is a function such that  $\phi(1_Q)=1_{Q'}$
Morphism of pointed sets induce morphisms of the associated pointed quandles
(cf. prop. \ref{prop:qu2}). 

\begin{prop} \label{prop:qumor3}
Let $Q,Q'$ be pointed sets and let $\phi:Q\rightarrow Q'$ be a pointed set morphism. 
Then, $\phi$ is a morphism of the pointed quandles $Q$, $Q'$.  
\end{prop}

\noindent
{\it Proof}. From \eqref{qu6}, we have $\phi(a\varrhd b)=\phi(b)=\phi(a)\varrhd'\phi(b)$
for $a, b\in  Q$ so that $\phi$ fulfills \eqref{qumor1}. 
The validity of \eqref{qumor2} is obvious, since $\phi$ is a pointed set morphism. 
\hfill $\Box$

If we denote $Q$, $Q'$ as $\Triv(Q)$, $\Triv(Q')$ to emphasize the quandle structure they are endowed with,
then the morphism $\phi$ gets denoted as $\Triv(\phi)$. The following proposition holds.

\begin{prop}
$\Triv$ is a functor from the category $\mathsans{PtSet}$ of pointed sets and pointed set morphisms 
to the category $\mathsans{PtQ}$ of pointed quandles and pointed quandle morphisms. 
\end{prop}

\noindent
{\it Proof}. Indeed, compositions and identities are obviously preserved by $\Triv$.
\hfill $\Box$


\subsection{\normalsize \textcolor{blue}{Quandle crossed modules}}\label{sec:qucros}

\hspace{.5cm} Just as crossed modules are generalization of groups involving two groups 
with certain structure maps, quandle crossed modules are generalizations
of quandles involving two quandles with additional structure maps.

\begin{defi} \label{def:qucros1}
A quandle crossed module consists of two quandles $Q$, $R$, 
a quandle morphism $\alpha:R\rightarrow Q$, called quandle target map,  
and a mapping $\varrhd:$ $Q\times R\rightarrow R$,
called quandle action, with the following properties.
\begin{subequations}
\begin{align}
&a\varrhd(b\varrhd A)=(a\varrhd b)\varrhd(a\varrhd A),
\vphantom{\Big]}
\label{qucros1}
\\
&a\varrhd(A\varrhd B)=(a\varrhd A)\varrhd(a\varrhd B),
\vphantom{\Big]}
\label{qucros2}
\\
&\alpha(a\varrhd A)=a\varrhd \alpha(A),
\vphantom{\Big]}
\label{qucros3}
\\
&\alpha(A)\varrhd B=A\varrhd B
\vphantom{\Big]}
\label{qucros4}
\end{align}
\end{subequations}
for arbitrary $a,b\in Q$, $A,B\in R$. Moreover, it is required that the equation 
$a\varrhd B=A$ has a unique solution $B\in R$ for any $a\in Q$, $A\in R$. 
\end{defi}
Since we use consistently lower case letters for the elements of $Q$ and upper case letters 
for those of $R$, it is clear from the context whether the symbol $\varrhd$ denotes the 
quandle pairing of $Q$ or that of $R$ or the quandle action of $Q$ on $R$. 
\eqref{qucros1} is the defining property of a quandle action. \eqref{qucros2} states that, for any $a\in Q$,
$a\varrhd\cdot:R\rightarrow R$ is a quandle automorphism of $R$. \eqref{qucros3}, \eqref{qucros4} are the
quandle crossed module relations, \eqref{qucros4} being the quandle version of the Peiffer identity.

\begin{defi} \label{def:qucros2}
A quandle crossed module $(Q,R)$ is said pointed if both $Q$, $R$ are pointed quandles, $\alpha$ is a morphism
of pointed quandles and further 
\begin{subequations}
\begin{align}
&1_Q\varrhd A=A,
\vphantom{\Big]}
\label{qucros5}
\\
&a\varrhd 1_R=1_R
\vphantom{\Big]}
\label{qucros6}
\end{align}
\end{subequations}
for any $a\in Q$, $A\in R$. 
\end{defi}

\begin{defi}  \label{def:qucros3}
An augmentation of a quandle crossed module $(Q,R)$ is a map $\varsucc:R\times Q\rightarrow Q$
satisfying the relation
\begin{equation}
a\varrhd(A\varsucc b)=(a\varrhd A)\varsucc(a\varrhd b)
\label{qucros7}
\end{equation}
and with the property \pagebreak 
that the equation $A\varsucc b=a$ has a unique solution $b\in Q$ for any
$a\in Q$ and $A\in R$. When $(Q,R)$ is pointed, we require in addition that
\begin{subequations}
\begin{align}
&A\varsucc 1_Q=\alpha(A),
\vphantom{\Big]}
\label{qucros8}
\\
&1_R\varsucc a=a
\vphantom{\Big]}
\label{qucros9}
\end{align}
\end{subequations}
for any $a\in Q$ and $A\in R$.
A quandle crossed module endowed with an augmentation is said augmented.
\end{defi}

Just as an ordinary group yields canonically an ordinary pointed quandle (cf. prop. \ref{prop:qu1}), 
a crossed module yields canonically an augmented pointed quandle crossed module. 

\begin{prop} \label{prop:qucros1} 
Let $(Q,R)$ be a crossed module with target map $t$ and action map $m$. 
View $Q$, $R$ as the pointed quandles of the underlying groups. 
Let $\alpha:R$ $\rightarrow Q$ and $\varrhd: Q\times R\rightarrow R$ be defined by 
setting
\begin{align}
&\alpha(A)=t(A),
\vphantom{\Big]}
\label{qucros10/0}
\\
&a\varrhd A=m(a)(A)
\vphantom{\Big]}
\label{qucros10}
\end{align}
with $a\in Q$ and $A\in R$. Let further $\varsucc: R\times Q\rightarrow Q$ be defined by
\begin{equation}
A\varsucc a=at(A)
\label{qucros11}
\end{equation}
with $a\in Q$ and $A\in R$.  
Then, $(Q,R)$ is an augmented pointed quandle crossed module.
\end{prop}

\noindent
{\it Proof}. 
Recall that the quandle operations of $Q$ and $R$ 
are both defined by relation \eqref{qu5}. By \eqref{qucros10/0}, on account of prop. \ref{prop:qumor2},
as $t$ is a group morphism, $\alpha$ is a pointed quandle morphism. 
\eqref{qucros1}, \eqref{qucros2}, \eqref{qucros3} and \eqref{qucros4} 
hold by \eqref{qu5}, \eqref{qucros10/0}, \eqref{qucros10}
and the basic crossed module identities 
$m(ab)(A)$ $=m(a)(m(b)(A))$, $m(a)(AB)=m(a)(A)m(a)(B)$, 
$t(m(a)(A))=at(A)a^{-1}$ and $m(t(A))(B)=ABA^{-1}$,
$a,b\in Q$, $A,B\in R$, respectively. 
\eqref{qucros5} and \eqref{qucros6} are an immediate consequence  
of \eqref{qucros10} and the identities $m(1_Q)(A)=A$ and $m(a)(1_R)=1_R$,
$a\in Q$, $A\in R$. 
\eqref{qucros7} holds by \eqref{qu5}, \eqref{qucros10}, \eqref{qucros11} and again the identity
$at(A)a^{-1}=t(m(a)(A))$, $a\in Q$, $A\in R$. Finally, \eqref{qucros8} and \eqref{qucros9}
follow readily from \eqref{qucros10/0}, \eqref{qucros11} and the
identity $t(1_R)=1_Q$. \hfill $\Box$

\noindent
%
This quandle crossed module is called the conjugation quandle crossed module 
of the crossed module $(Q,R)$ and is denoted by $\Conj(Q,R)$, 
when one wants to emphasize its quandle theoretic structure.

As there exists a pointed quandle canonically associated to a pointed set
(cf. prop. \ref{prop:qu2}), there is an augmented pointed quandle crossed module 
associated to a pair of pointed sets. 

\begin{prop} \label{prop:qucros2} 
Let $Q$, $R$ be pointed sets. 
View $Q$, $R$ as the pointed quandles of the corresponding sets. 
Let $\alpha:R\rightarrow Q$ and $\varrhd: Q\times R\rightarrow R$ be defined by 
\begin{align}
&\alpha(A)=1_Q,
\vphantom{\Big]}
\label{qucros12/0}
\\
&a\varrhd A=A
\vphantom{\Big]}
\label{qucros12}
\end{align}
with $a\in Q$ and $A\in R$. Let further $\varsucc: R\times Q\rightarrow Q$ be defined by
\begin{equation}
A\varsucc a=a 
\label{qucros13}
\end{equation}
with $a\in Q$ and $A\in R$. Then, $(Q,R)$ is an augmented pointed quandle crossed module.
\end{prop}

\noindent
{\it Proof}. Recall that the quandle operations of $Q$ and $R$ 
are both defined by relation \eqref{qu6}. 
By \eqref{qucros12/0}, it is clear that $\alpha$ is a pointed quandle morphism.
\eqref{qucros1}, \eqref{qucros2}, 
\eqref{qucros3} and \eqref{qucros4} hold trivially by \eqref{qu6},
\eqref{qucros12/0}, \eqref{qucros12}. \eqref{qucros5} and \eqref{qucros6} 
are an immediate consequence of \eqref{qucros12}. \eqref{qucros7} holds by \eqref{qu6}
and \eqref{qucros13}. Finally, \eqref{qucros8} and \eqref{qucros9}
are follow trivially from \eqref{qucros12/0} and \eqref{qucros13}. 
\hfill $\Box$

\noindent
This quandle crossed module is called the trivial quandle crossed module 
of the pointed set pair $(Q,R)$ and is denoted by $\Triv(Q,R)$, 
when one wishes to emphasize its having a quandle structure.


\subsection{\normalsize \textcolor{blue}{Morphisms of quandle crossed modules}}\label{sec:qucromor}

\hspace{.5cm} Quandle crossed module morphisms are mappings between quandles crossed modules 
respecting their quandle structure. 

\begin{defi}
Let $(Q,R)$, $(Q',R')$ be quandle crossed modules. A morphism $(\phi,\psi):(Q,R)\rightarrow (Q',R')$
consists of a pair of quandle morphisms $\phi:Q\rightarrow Q'$, $\psi:R\rightarrow R'$ 
such that \hphantom{xxxxxxxxxxxxxxxx}
\begin{align}
&\phi(\alpha(A))=\alpha'(\psi(A))
\vphantom{\Big[}
\label{qucromor1}
\\
&\psi(a\varrhd A)=\phi(a)\varrhd\!{}'\,\psi(A)
\vphantom{\Big[}
\label{qucromor3}
\end{align}
for $a\in Q$ and $A\in R$. If $(Q,R)$, $(Q',R')$ are both pointed, it is further required that
$\phi$, $\psi$ be pointed quandle morphisms. If $(Q,R)$, $(Q',R')$ are both endowed with a
augmentation, it is also required that
\begin{equation}
\phi(A\varsucc a)=\psi(A)\varsucc\!{}'\,\phi(a)
\label{qucromor4}
\end{equation}
for $a\in Q$ and $A\in R$.
\end{defi}


Like quandles, quandle crossed modules and their 
morphisms form a category.

\begin{prop}
The quandle crossed modules and quandle crossed module morphism form a category. 
The pointed, augmented and pointed and augmented quandle crossed modules and crossed 
module morphisms constitute subcategories of this category. 
\end{prop}

\noindent
{\it Proof}. From \eqref{qucromor1}, \eqref{qucromor3}, it is easily shown that the composition 
$(\phi'\circ\phi,\psi'\circ\psi):
(Q,R)\rightarrow (Q'',R'')$ of two quandle crossed module 
morphisms $(\phi,\psi):(Q,R)\rightarrow (Q',R')$, $(\phi',\psi'):(Q',R')\rightarrow (Q'',R'')$
the identity $(\id_Q,\id_R)$ of a quandle crossed module are quandle
crossed module morphisms. Therefore, quandles and quandle morphisms
constitute a subcategory of the category $\mathsans{Set}\times\mathrm{Set}$, hence a category itself. 
Similar conclusions are reached in the pointed, augmented and pointed and augmented 
cases.
\hfill $\Box$

We shall denote by $\mathsans{QCM}$ (resp. $\mathsans{PtQCM}$, $\mathsans{AuQCM}$, $\mathsans{AuPtQCM}$),  
the category of ordinary (resp. pointed, augmented, pointed and augmented) quandle crossed modules 
and quandle crossed module morphisms.

\begin{prop}
Let $(Q,R)$ be a (pointed, augmented, pointed and augmented) quandle crossed module. For $a\in Q$, 
let $\iota_a:Q\rightarrow Q$ and $\mu_a:R\rightarrow R$ be the mappings defined by the expressions
\begin{subequations}
\label{qucromor5,6}
\begin{align}
&\iota_a(b)=a\varrhd b,
\vphantom{\Big]}
\label{qucromor5}
\\
&\mu_a(B)=a\varrhd B
\vphantom{\Big]}
\label{qucromor6}
\end{align}
\end{subequations}
with $b\in Q$ and $B\in R$. Then, $(\iota_a,\mu_a)$   
is a (pointed, augmented, pointed and augmented) quandle crossed module automorphism.
\end{prop}

\noindent
{\it Proof}. First of all, we note that the maps $\iota_a$, $\mu_a$ are invertible by one of the defining 
properties of a quandle crossed module structure (cf. defs. \ref{def:qu1}, \ref{def:qucros1}). 
By \eqref{qu2}, \eqref{qucros2}, further, we have 
$\iota_a(b\varrhd c)=a\varrhd (b\varrhd c)=(a\varrhd b)\varrhd(a\varrhd c)$ $=\iota_a(b)\varrhd\iota_a(c)$,
$\mu_a(B\varrhd C)=a\varrhd (B\varrhd C)=(a\varrhd B)\varrhd(a\varrhd C)
=\mu_a(B)\varrhd\mu_a(C)$, showing that $\iota_a$, $\mu_a$ are automorphisms of the
quandles $Q$ and $R$, respectively. Next, by \eqref{qucros3}, \eqref{qucros1}, we have 
$\iota_a(\alpha(B))=a\varrhd\alpha(B)=\alpha(a\varrhd B)=\alpha(\mu_a(B))$,
$\mu_a(b\varrhd B)=a\varrhd(b\varrhd B)=(a\varrhd b)\varrhd(a\varrhd B)
=\iota_a(b)\varrhd \mu_a(B)$ with $b\in Q$, $B\in R$, verifying
\eqref{qucromor1}, \eqref{qucromor3} and showing 
$(\iota_a,\mu_a):(Q,R)\rightarrow(Q,R)$ is a quandle crossed module 
automorphism of $(Q,R)$.
Let $(Q,R)$ be pointed. By \eqref{qu3}, \eqref{qucros6}, we have 
$\iota_a(1_Q)=a\varrhd 1_Q=1_Q$, $\mu_a(1_R)=a\varrhd 1_R=1_R$, verifying 
that the quandle automorphisms $\iota_a$, $\mu_a$ are both pointed
as required. 
Let $(Q,R)$ be augmented. Then, by \eqref{qucros7}, 
$\iota_a(B\varsucc b)=a\varrhd(B\varsucc b)=(a\varrhd B)\varsucc(a\varrhd b)
=\mu_a(B)\varsucc\iota_a(b)$ proving \eqref{qucros7} as required. 
\hfill $\Box$

\begin{defi}
Let $(Q,R)$ be a (pointed, augmented, pointed and augmented) quandle crossed module. 
A (pointed, augmented, pointed and augmented) quandle crossed module automorphism of $(Q,R)$ 
is called inner if it is of the form $(\iota_a,\mu_a)$ for some $a\in Q$.  
\end{defi}

The subgroup of the (pointed, augmented, pointed and augmented) quandle crossed module automorphism group 
$\Aut(Q,R)$ of $(Q,R)$ generated by  
the inner elements  and their inverses will be denoted by $\Inn(Q,R)$.

\begin{prop}
Let $(Q,R)$ be a (pointed, augmented, pointed and augmented) quandle crossed module. 
Then, $\Inn(Q,R)$ is a normal subgroup of $\Aut(Q,R)$.  
\end{prop}

\noindent
{\it Proof}. Let $a\in Q$ and $(\phi,\psi)$ be an arbitrary quandle crossed module automorphism of $(Q,R)$. 
Using \eqref{qumor1}, \eqref{qucromor3}, we find 
that $\phi\circ\iota_a\circ \phi^{-1}(b)=\phi(a\varrhd\phi^{-1}(b))=\phi(a)\varrhd b=\iota_{\phi(a)}(b)$, 
$\psi\circ\mu_a\circ \psi^{-1}(B)=\psi(a\varrhd \psi^{-1}(B))=\phi(a)$ $\varrhd B=\mu_{\phi(a)}(B)$. 
for $b\in Q$, $B\in R$. Since  the subgroup $\Inn(Q,R)$ of $\Aut(Q,R)$ is generated by the automorphisms 
$(\iota_a,\mu_a)$ and their inverses, $\Inn(Q,R)$ is invariant, hence normal in $\Aut(Q,R)$. \hfill $\Box$

Morphisms of crossed module induce morphisms of the associated conjugation 
quandle crossed modules (cf. prop. \ref{prop:qucros1}). 

\begin{prop}
Let $(Q,R)$, $(Q',R')$ be crossed modules and $(\phi,\psi):(Q,R)$ $\rightarrow (Q',R')$
be a crossed module morphism. View $(Q,R)$, $(Q',R')$ as augmented pointed quandle crossed 
modules. Then, $(\phi,\psi)$ is a morphism of augmented pointed quandle crossed modules. 
\end{prop}

\noindent
{\it Proof}. Recall that 
a morphism $(\phi,\psi):(Q,R)\rightarrow (Q',R')$ of crossed modules
is a pair of group morphisms $\phi:Q\rightarrow Q'$, $\psi:R\rightarrow R'$
such that $\phi(t(A))=t'(\psi(A))$ and $\psi(m(a)(A))=m'(\phi(a))(\psi(A))$
with $a\in Q$, $A\in R$. By prop. \ref{prop:qumor2}, $\phi:Q\rightarrow Q'$, $\psi:R\rightarrow R'$
are then morphisms of pointed quandles. 
From relations \eqref{qucros10/0}, \eqref{qucros10} and \eqref{qucros11}, we have 
$\phi(\alpha(A))=\phi(t(A))=t'(\psi(A))=\alpha'(\psi(A))$, 
$\psi(a\varrhd A)=\psi(m(a)(A))=m'(\phi(a))(\psi(A))=\phi(a)\varrhd\!{}'\,\psi(A)$ and 
$\phi(A\varsucc a)=\phi(a)\phi(t(A))=\phi(a)t'(\psi(A))=\psi(A)\varsucc\!{}'\,\phi(a)$
for $a\in Q$, $A\in R$ verifying the axioms \eqref{qucromor1}, \eqref{qucromor3} as well as 
\eqref{qucromor4}. \hfill $\Box$

If we denote $(Q,R)$, $(Q',R')$ as $\Conj(Q,R)$, $\Conj(Q',R')$ to emphasize their being endowed with a quandle
crossed module structure, then the morphism $(\phi,\psi)$ gets denoted as $\Conj(\phi,\psi)$. 
The following proposition holds.

\begin{prop}
$\Conj$ is a functor from the category $\mathsans{CM}$ of crossed modules and crossed module morphisms 
to the category $\mathsans{AuPtQCM}$ of augmented pointed quandle crossed modules and quandle crossed module
morphisms. 
\end{prop}

\noindent
{\it Proof}. Indeed, compositions and identities are obviously preserved by $\Conj$.
\hfill $\Box$

An analogous result holds also for trivial quandle crossed modules (cf. prop. \ref{prop:qucros2}). 

\begin{prop}
Let $Q$, $R$, $Q'$, $R'$ be pointed sets. Let $\phi:Q\rightarrow Q'$, $\psi:R\rightarrow R'$ be two pointed maps.
View $(Q,R)$, $(Q',R')$ as augmented pointed quandle crossed modules. Then, 
$(\phi,\psi):(Q,R)\rightarrow (Q',R')$ is a morphism of augmented pointed quandle crossed modules. 
\end{prop}

\noindent
{\it Proof}. By prop. \ref{prop:qumor3}, $\phi:Q\rightarrow Q'$, $\psi:R\rightarrow R'$ are pointed quandle
morphisms. From \eqref{qucros12/0}, \eqref{qucros12} and \eqref{qucros13}, we have that 
$\phi(\alpha(A))=\phi(1_Q)=1_{Q'}$ \linebreak $=\alpha'(\psi(A))$, $\psi(a\varrhd A)=\psi(A)=\phi(a)\varrhd\!{}'\,\psi(A)$
and $\phi(A\varsucc a)=\phi(a)=\psi(A)$ $\varsucc\!{}'\,\phi(a)$
for $a\in Q$, $A\in R$, verifying \eqref{qucromor1}, \eqref{qucromor3}
 as well as \eqref{qucromor4}. \hfill $\Box$

If we denote $(Q,R)$, $(Q',R')$ as $\Triv(Q,R)$, $\Triv(Q',R')$ to emphasize the quandle crossed module structure they are endowed with,
then the morphism $(\phi,\psi)$ gets denoted as $\Triv(\phi,\psi)$. The following proposition holds.

\begin{prop}
$\Triv$ is a functor from the category $\mathsans{PtSet}\times \mathsans{PtSet}$ of ordered pairs of pointed sets and pointed 
set morphisms to the category $\mathsans{AuPtQCM}$ of augmented pointed quandle crossed modules and quandle crossed module
morphisms. 
\end{prop}

\noindent
{\it Proof}. Indeed, compositions and identities are obviously preserved by $\Triv$.
\hfill $\Box$

\vfil\eject

\section{\normalsize \textcolor{blue}{Higher gauge quandle and holonomy invariants}}\label{sec:holoinv}

\hspace{.5cm} 
In this section, we shall show how the abstract quandle theory expounded in the previous section can 
be fruitfully applied to define and construct invariant traces over crossed modules. 
In turn these can be used to build holonomy invariants in strict higher gauge theory
for reasons explained in subsect. \ref{sec:scope} of the introduction, which we shall not repeat. 
In subsect. \ref{sec:hitr}, we describe invariant traces abstractly as morphisms
of the conjugation quandle of the relevant gauge group or crossed module 
into a trivial numerical quandle or quandle crossed module, respectively. 
In subsect. \ref{sec:hirep}, we illustrate a natural construction of invariant traces 
using representations of the gauge group or crossed module.


\subsection{\normalsize \textcolor{blue}{Traces over crossed modules}}\label{sec:hitr}

\hspace{.5cm} The definition of holonomy invariants requires a notion of trace. 
Here, we present first an axiomatization of the ordinary group trace  based
on quandle theory and then we propose a natural extension to crossed module 
trace relying on quandle crossed module theory. 

Let $G$ be a group. 

\begin{defi} \label{def:hitr1}
A trace over $G$ is a mapping $\tr:G\rightarrow \mathbb{C}$ such that 
\begin{equation}
\tr(aba^{-1})=\tr(b)
\label{hitr1}
\end{equation}
for arbitrary $a,b\in G$.
\end{defi}
This is the only property of any trace which really matters if the gauge invariance 
of holonomy invariants is to be ensured. 

\begin{defi} \label{def:hitr2}
The characteristics value of a trace $\tr$ over $G$ is 
\begin{equation}
\kappa_{\tr}=\tr(1_G).
\label{hitr2}
\end{equation}
\end{defi}
We observe that if $\tr$ is a trace over $G$ so is $z\cdot \tr$ for any complex number
$z\in \mathbb{C}$. 
So, we can normalize the trace by requiring that
$\kappa_{\tr}\in\mathbb{Z}$ is integer. 

The above can be aptly rephrased using 
quandle theory
(cf. sect. \ref{sec:quandle}). 
With the group $G$, there is associated  a pointed quandle, 
the conjugation quandle $\Conj(G)$ of $G$.
This is equal to $G$ as a set, its quandle operation is defined by 
\begin{equation}
a\varrhd b=aba^{-1}
\label{hitr3}
\end{equation}
with $a,b\in G$ and its neutral element is $1_G$ (cf. subsect. \ref{sec:qu}). 
Likewise, with the set $\mathbb{C}$ pointed by $\kappa_{\tr}$, 
there is associated a pointed quandle, the trivial quandle $\Triv(\mathbb{C})$ of $\mathbb{C}$. 
This is equal to $\mathbb{C}$ as a set, its quandle operation is defined by 
\begin{equation}
w\varrhd z=z
\label{hitr4}
\end{equation}
with $w,z\in\mathbb{C}$ and its neutral element is $\kappa_{\tr}$ (cf. subsect. \ref{sec:qu}).
A trace over $G$ can now be viewed as quandle morphism of these two quandles
(cf. subsect. \ref{sec:qumor}). 

\begin{prop} \label{prop:hitr1}
If $\tr$  is a trace over $G$, then $\tr$ is a pointed quandle morphism of $\Conj(G)$ into 
$\Triv(\mathbb{C})$. 
\end{prop}

\vspace{1mm} 
\noindent
{\it Proof}. In fact, relation \eqref{hitr1} can be written as
\begin{equation}
\tr(a\varrhd b)=\tr(b)=\tr(a)\varrhd\tr(b)
\label{hitr5}
\end{equation}
(cf. eq. \eqref{qumor1}) and \eqref{hitr2} is the statement of the pointedness of $\tr$
(cf. eq. \eqref{qumor2}). \hfill $\Box$

Our quandle theoretic formulation of the notion of trace over a group 
points to a possible viable definition of trace over a crossed module.

Let $(G,H)$ be a crossed module with target and action maps $t$ and $m$, respectively. 

\begin{defi} \label{def:hitr3}
A trace over $(G,H)$ is a pair $(\tr_b,\tr_f)$ of mappings $\tr_b:G\rightarrow \mathbb{C}$, 
$\tr_f:H\rightarrow \mathbb{C}$ satisfying the relations 
\begin{subequations}
\begin{align}
&\tr_b(aba^{-1}t(A))=\tr_b(b),
\vphantom{\Big]}
\label{hitr6}
\\
&\tr_f(m(a)(B))=\tr_f(B)
\vphantom{\Big]}
\label{hitr7}
\end{align}
\end{subequations}
for arbitrary $a,b\in G$ and $A,B\in H$. 
\end{defi}
The above are the minimal properties required to any crossed module trace to ensure the gauge invariance 
of higher holonomy invariants. 

We observe here that \eqref{hitr6}, \eqref{hitr7} imply that 
$\tr_b$, $\tr_f$ are traces over the groups $G$ and $H$, respectively, in the second case thanks to the 
Peiffer identity. 

\begin{defi} \label{def:hitr4}
The characteristics values of a trace $(\tr_b,\tr_f)$ over $(G,H)$ are 
\begin{subequations}
\begin{align}
&\kappa_{\tr b}=\tr_b(1_G),
\vphantom{\Big]}
\label{hitr8}
\\
&\kappa_{\tr f}=\tr_f(1_H).
\vphantom{\Big]}
\label{hitr9}
\end{align}
\end{subequations}
\end{defi}
From the definition,  if $(\tr_b,\tr_f)$ is a trace over $(G,H)$ so is $(z\cdot \tr_b,w\cdot\tr_f)$ for any complex numbers
$z,w\in \mathbb{C}$. Using this property, we can normalize the trace by requiring that
$\kappa_{\tr b},\kappa_{\tr f}\in\mathbb{Z}$ are integer. 

As in the ordinary case, one can be elegantly rephrase the above using quandle theory
(cf. sect. \ref{sec:quandle}). 
With $(G,H)$, there is associated an augmented pointed quandle crossed module, 
the conjugation quandle crossed module $\Conj(G,H)$ of $(G,H)$.
This consists of the pointed quandles $G$, $H$, with target map 
\begin{equation}
\alpha(A)=t(A),
\label{hitr10}
\end{equation} 
quandle action \hphantom{xxxxxxxxxxxxxxxx}
\begin{equation}
a\varrhd A=m(a)(A)
\label{hitr11}
\end{equation}
and augmentation \hphantom{xxxxxxxxxxxxxxxx}
\begin{equation}
A\varsucc a=at(A),
\label{hitr12}
\end{equation}
where $a\in G$ and $A\in H$ (cf. subsect. \ref{sec:qucros}). 
Additionally, one can construct an augmented pointed quandle crossed module
from two copies $\mathbb{C}_b$, $\mathbb{C}_f$ of $\mathbb{C}$ pointed by 
$\kappa_{\tr b}$, $\kappa_{\tr f}$ respectively, the trivial quandle crossed module $\Triv(\mathbb{C}_b,\mathbb{C}_f)$
of the pair $(\mathbb{C}_b,\mathbb{C}_f)$.
This consists of the pointed quandles $\mathbb{C}_b$, $\mathbb{C}_f$, 
the target map 
\begin{equation}
k(Z)=\kappa_{\tr b},
\label{hitr13}
\end{equation}
quandle action \hphantom{xxxxxxxxxxxxxxxxxxxx}
\begin{equation}
z\varrhd Z=Z
\label{hitr14}
\end{equation}
and augmentation \hphantom{xxxxxxxxxxxxxxxxxx}
\begin{equation}
Z\varsucc z=z,
\label{hitr15}
\end{equation}
where $z\in \mathbb{C}_b$ and $Z\in \mathbb{C}_f$ (cf. subsect. \ref{sec:qucros}). 
A trace over $(G,H)$ can now be viewed as augmented pointed quandle crossed 
module morphism of these two quandles (cf. subsect. \ref{sec:qucromor}). 

\begin{prop} \label{prop:hitr2}
If $(\tr_b,\tr_f)$ is a trace over $(G,H)$, then $(\tr_b,\tr_f)$ 
is an augmented pointed quandle crossed module morphism of $\Conj(G,H)$ into $\Triv(\mathbb{C}_b,\mathbb{C}_f)$. 
\end{prop}

\noindent{\it Proof}. Albeit straightforward, the proof illustrates how quandle theory is well suited 
to the description of the invariance properties of traces.

By \eqref{hitr6} with $A=1_H$, we have 
\begin{equation}
\tr_b(a \varrhd  b)
=\tr_b(aba^{-1})=\tr_b(b)=\tr_b(a)\varrhd \tr_b(b)
\label{}
\end{equation}
for $a,b\in G$, showing that $\tr_b$ is a morphism of the quandle $G$ into $\mathbb{C}_b$
(cf. eq. \eqref{qumor1}). 
Further, as $\tr_b(1_G)=\kappa_{\tr b}$, $\tr_b$ is pointed.
By \eqref{hitr7} with $a$, $A$ replaced by $t(A)$, $B$ respectively and the Peiffer identity, we have 
\begin{align}
&\tr_f(A \varrhd  B)=\tr_f(ABA^{-1})
=\tr_f(m(t(A))(B))
\label{}
\\
&\hspace{5cm}=\tr_f(B)=\tr_f(A)\varrhd \tr_f(B)
\nonumber
\end{align}
for $A,B\in H$, showing that $\tr_f$ is a morphism of the quandle $H$ into $\mathbb{C}_f$ (cf. eq. \eqref{qumor1}). 
Further, as $\tr_f(1_H)=\kappa_{\tr f}$, $\tr_f$ is pointed.
By \eqref{hitr6}, \eqref{hitr10} and \eqref{hitr13}, we also have 
\begin{equation}
\tr_b(\alpha(A))=
\tr_b(t(A))=\tr_b(1_G)=\kappa_{\tr b}=k(\tr_f(A))
\label{}
\end{equation}
for $A\in H$, verifying condition \eqref{qucromor1}. As \eqref{hitr7} can be written as 
\begin{equation}
\tr_f(a\varrhd B)=\tr_f(B)=\tr_b(a)\varrhd\tr_f(B)
\label{}
\end{equation}
with $a\in G$ and $B\in H$ by \eqref{hitr11}, \eqref{qucromor3} is satisfied. Finally, 
from \eqref{hitr6} with $a$ replaced by $1_G$, we find that 
\begin{equation}
\tr_b(A\varsucc b)=\tr_b(bt(A))=\tr_b(b)=\tr_f(A)\varsucc\tr_b(b)
\label{}
\end{equation}
with $b\in G$ and $A\in H$, so that also \eqref{qucromor4} is also satisfied. 
All the 
required condition being fulfilled, $(\tr_b,\tr_f)$  is an augmented pointed quandle crossed morphism
of $\Conj(G,H)$ into $\Triv(\mathbb{C}_b,\mathbb{C}_f)$. \hfill $\Box$


\subsection{\normalsize \textcolor{blue}{Representations of crossed modules and traces}}\label{sec:hirep}

\hspace{.5cm} In this subsection, we shall illustrate a general scheme for 
the construction of traces over a Lie crossed module. 

Traces on groups are easy to construct. 
Let $G$ be a Lie group and $R:G\rightarrow \GL(X)$ be a representation of $G$ in 
the vector space $X$.

\begin{defi} \label{def:hirep0}
For $a\in G$, set
\begin{equation}
\tr_R(a)=\tr_X(R(a))
\label{hirep-1}
\end{equation}
\end{defi}
In common parlance, $\tr_R$ is the character of the representation $R$.

\begin{prop} \label{prop:hirep-1}
$\tr_R$ is a trace over $G$
\end{prop}

\noindent
{\it Proof}. The proof reduces to the verification $\tr_R$ satisfies condition
\eqref{hitr1}, which is trivial. \hfill $\Box$

\noindent 
The characteristic value of the trace $\tr_R$ (cf. def. \ref{def:hitr2})
is the dimension of the representation $R$

\begin{prop} \label{prop:hirep0}
One has 
\begin{equation}
\kappa_{\tr R}=\dim X.
\label{hirep0}
\end{equation}
\end{prop}

\noindent
{\it Proof}. This follows trivially from 
the defining relation \eqref{hitr2} and \eqref{hirep0}. 
\hfill $\Box$

Constructing traces over Lie crossed modules turns out to be not so straightforward. 

\begin{defi} \label{def:hirep1}
Let $X,Y$ be vector spaces on the same field. A Lie crossed module $(G,H)$ is said of 
type $(X,Y)$ if $G\subset \GL(X)$, $H\subset \GL(Y)$.  
\end{defi}

\begin{defi} \label{def:hirep2}
Let $X,Y$ be complex vector spaces equipped with an Hermitian inner product. A Lie crossed module $(G,H)$ 
of type $(X,Y)$ is said unitary if $G\subset \UU(X)$, $H\subset \UU(Y)$.  
\end{defi}

Let $(G,H)$ be a Lie crossed module with target and action maps $t$ and $m$
and $(X,Y)$ be a pair of vector spaces on the same field.

\begin{defi} \label{def:hirep3}
A representation of $(G,H)$ on $(X,Y)$ is a Lie crossed module morphism $(R,S):(G,H)\rightarrow (G',H')$,
where $(G',H')$ is a Lie crossed module of type $(X,Y)$. The representation is said unitary
if $X, Y$ are complex vector spaces with Hermitian inner product and $(G',H')$ is unitary. 
\end{defi}

\noindent
Explicitly, the representation consists of two group morphisms $R:G\rightarrow G'$, $S:H\rightarrow H'$
such that 
\begin{subequations}
\label{hirep1,2}
\begin{align}
&R(t(A))=t'(S(A)),
\vphantom{\Big]}
\label{hirep1}
\\
&S(m(a)(A))=m'(R(a))(S(A))
\vphantom{\Big]}
\label{hirep2}
\end{align}
\end{subequations}
with $a\in G$, $A\in H$. 

From now on, we assume that {\it $G$, $H$ are compact Lie groups}.
With no loss of generality, we can then take 
$(R,S)$ to be a unitary representation of $(G,H)$, since for compact groups
every representation is equivalent to a unitary one. 

Let $\mu_G$, $\mu_H$ be the bi-invariant Haar measures of $G$, $H$
normalized so that $\vol(G)=\vol(H)=1$ \cite{Haar:1965hm}.
By a standard group averaging, we  define the following mappings. 
\begin{defi} \label{def:hirep4}
For $a\in G$ and $A\in H$, set
\begin{subequations}
\label{hirep3,4}
\begin{align}
&\tr_{R,Sb}(a)=\int_Hd\mu_H(X)\tr_X(R(at(X))),
\vphantom{\Big]}
\label{hirep3}
\\
&\tr_{R,Sf}(A)=\int_Gd\mu_G(x)\tr_Y(S(m(x)(A))).
\vphantom{\Big]}
\label{hirep4}
\end{align}
\end{subequations}
\end{defi}

\noindent
These constitute a crossed  module trace (cf. def. \ref{def:hitr4}).

\begin{prop} \label{prop:hirep1}
$(\tr_{R,Sb},\tr_{R,Sf})$ is a trace over $(G,H)$.
\end{prop}

\noindent
{\it Proof}. As a preliminary result let us prove that, 
for any continuous function $f:H\rightarrow \mathbb{C}$ and $a\in G$, 
\begin{equation}
\int_Hd\mu_H(X)f(m(a)(X))=\int_Hd\mu_H(X)f(X).
\label{hirepa1}
\end{equation}
Consider the algebra $C(H)$ of continuous functions on the topological group $H$ 
and the linear functional $\mathcal{F}:C(H)\rightarrow \mathbb{C}$ defined by
\begin{equation}
\mathcal{F}(g)=\int_Gd\mu_G(x)\int_Hd\mu_H(X) g(m(x)(X))
\label{hirepa2}
\end{equation}
with $g\in C(H)$.
For $A\in H$, let $L_A$ denote the left translation by $A$ in $H$, so that $L_A(B)=AB$ for $B\in H$.
We then have 
\begin{align}
\mathcal{F}(g\circ L_A)&=\int_Gd\mu_G(x)\int_Hd\mu_H(X) g(Am(x)(X))
\vphantom{\Big]}
\label{hirepa3}
\\
&=\int_Gd\mu_G(x)\int_Hd\mu_H(X) g(m(x)(m(x)^{-1}(A)X))
\vphantom{\Big]}
\nonumber
\\
&=\int_Gd\mu_G(x)\int_Hd\mu_H(X) g(m(x)(X))=\mathcal{F}(g)
\vphantom{\Big]}
\nonumber
\end{align}
by the left invariance of the Haar measure of $\mu_H$. 
By the Riesz representation theorem and the uniqueness up to a constant of the Haar measure
of $H$, there is a positive constant $c$ such that 
\begin{equation}
\mathcal{F}(g)=c\int_Hd\mu_H(X)g(X)
\label{hirepa4}
\end{equation}
for $g\in C(H)$. By \eqref{hirepa1} and \eqref{hirepa5} with $g=1$, we have
\begin{equation}
c=c\vol(H)=\mathcal{F}(1)=\vol(G)\vol(H)=1.
\label{hirepa5}
\end{equation}
Inserting this value into \eqref{hirepa4}, we find that 
\begin{equation}
\mathcal{F}(g)=\int_Hd\mu_H(X)g(X).
\label{hirepa6}
\end{equation}
By virtue of \eqref{hirepa1} and \eqref{hirepa6}, for any continuous function $f:H\rightarrow \mathbb{C}$, 
\begin{align}
\int_Hd\mu_H(X)f(m(a)(X))
&=\mathcal{F}(f\circ m(a)) \hspace{4cm}
\vphantom{\Big]}
\label{hirepa7}
\\
&=\int_Gd\mu_G(x)\int_Hd\mu_H(X) f(m(a)(m(x)(X)))
\vphantom{\Big]}
\nonumber
\\
&=\int_Gd\mu_G(x)\int_Hd\mu_H(X) f(m(ax)(X))
\vphantom{\Big]}
\nonumber
\\
&=\int_Gd\mu_G(x)\int_Hd\mu_H(X) f(m(x)(X))
\vphantom{\Big]}
\nonumber
\\
&=\mathcal{F}(f)
\vphantom{\Big]}
\nonumber
\\
&=\int_Hd\mu_H(X) f(X),
\vphantom{\Big]}
\nonumber
\end{align}
where the left invariance of the Haar measure $\mu_G$ was exploited. 
\eqref{hirepa1} is therefore shown. 

We now can prove the proposition by verifying that conditions \eqref{hitr6}, \eqref{hitr7} hold.
Let $a, b\in G$ and $A\in H$. Then, by \eqref{hirepa1}  and the left invariance of $\mu_H$
\begin{align}
\tr_{R,Sb}(aba^{-1}t(A))
&=\int_Hd\mu_H(X)\tr_X(R(aba^{-1}t(A)t(X)))
\vphantom{\Big]}
\label{hirepa8}
\\
&=\int_Hd\mu_H(X)\tr_X(R(bt(m(a^{-1})(AX))))
\vphantom{\Big]}
\nonumber
\\
&=\int_Hd\mu_H(X)\tr_X(R(bt(m(a^{-1})(X))))
\vphantom{\Big]}
\nonumber
\\
&=\int_Hd\mu_H(X)\tr_X(R(bt(X)))=\tr_{R,Sb}(b),
\vphantom{\Big]}
\nonumber
\end{align}
verifying \eqref{hitr6}. Next, let $a\in G$ and $A\in H$. Then, 
by the right invariance of the Haar measure $\mu_G$,
\begin{align}
\tr_{R,Sf}(m(a)(A))&
=\int_Gd\mu_G(x)\tr_Y(S(m(x)(m(a)(A))))
\vphantom{\Big]}
\label{hirepa9}
\\
&=\int_Gd\mu_G(x)\tr_Y(S(m(xa)(A)))
\vphantom{\Big]}
\nonumber
\\
&=\int_Gd\mu_G(x)\tr_Y(S(m(x)(A)))=\tr_{R,Sf}(A),
\vphantom{\Big]}
\nonumber
\end{align}
verifying \eqref{hitr7}. \hfill $\Box$

The characteristic values of the crossed module trace $(\tr_{R,Sb},\tr_{R,Sf})$
are given by the following proposition, which is the counterpart of prop. \ref{prop:hirep0}
for crossed modules. 

\begin{prop} \label{prop:hirep2}
One has 
\begin{subequations}
\label{hirep5,6}
\begin{align}
\kappa_{\tr R,Sb}&=\int_Hd\mu_H(X)\tr_X(R(t(X))),
\vphantom{\Big]}
\label{hirep5}
\\
\kappa_{\tr R,Sf}&=\dim Y, 
\vphantom{\Big]}
\label{hirep6}
\end{align}
\end{subequations}
both numbers being integer.
\end{prop}

\noindent
{\it Proof}. \eqref{hirep5}, \eqref{hirep6} follow immediately form the defining relations
\eqref{hitr8}, \eqref{hitr9} and the \eqref{hirep3}, \eqref{hirep4}. 
The integrality of $\kappa_{\tr R,Sb}$ stems from noting that $X\rightarrow \tr_X(R(t(X)))$
is the character of the representation $R\circ t$ of $H$ and that the integral
\eqref{hirep5} equals the number of time the trivial representation of $H$ is contained
in the reduction of $R\circ t$ in irreducible representations, by the Peter--Weyl theorem. 
The integrality of $\kappa_{\tr R,Sf}$ is obvious. \hfill $\Box$

If $\kappa_{\tr R,Sb}\not=0$, it is possible to renormalize the trace $(\tr_b,\tr_f)$ as 
\begin{subequations}
\label{hirep7,8}
\begin{align}
&\overline{\tr}_{R,Sb}(a)=\frac{\dim X}{\kappa_{\tr R,Sb}}\tr_{R,Sb}(a),
\vphantom{\Big]}
\label{hirep7}
\\
&\overline{\tr}_{R,Sf}(A)=tr_{R,Sf}(A)
\vphantom{\Big]}
\label{hirep8}
\end{align}
\end{subequations}
with $a\in G$ and $A\in H$. 
The renormalized characteristic values are then
\begin{subequations}
\label{hirep9.10}
\begin{align}
\overline{\kappa}_{\tr R,Sb}&=\dim X,
\vphantom{\Big]}
\label{hirep9}
\\
\overline{\kappa}_{\tr R,Sf}&=\dim Y.
\vphantom{\Big]}
\label{hirep10}
\end{align}
\end{subequations}
This makes the analogy to the group theoretic case closer.


\subsection{\normalsize \textcolor{blue}{Generalized traces}}\label{sec:gentr}

\hspace{.5cm} The presentation of traces as morphisms of quandle structures of
subsect. \ref{sec:hitr} is amenable of a generalization discussed in
the present subsection.

Let $G$ be a group. Prop. \ref{prop:hitr1} suggests the following definition.

\begin{defi} \label{def:gentr1}
A trace $\tr$ over $G$ is a pointed quandle morphism of the 
conjugation quandle $G$ into some target pointed quandle $T$. 
\end{defi}

\noindent 
When $T$ is the trivial pointed quandle $\Triv(\mathbb{C})$, we recover the restricted 
notion of trace of def. \ref{def:hitr1}.

By \eqref{qumor1}, the condition that $\tr$ is a trace over $G$ entails that 
\begin{equation}
\tr(a\varrhd b)=\iota_{\tr(a)}(\tr(b))
\label{gentr1}
\end{equation}
with $a,b\in G$, where $\iota_x$ denotes the inner automorphism of $T$ associated with $x\in T$ (cf. eq. \eqref{qumor3}). Hence, 
for fixed $b\in G$, $\tr(a\varrhd b)$ with $a\in G$ varies in the orbit $\Orb_{\Inn}(\tr(b))$ of $\tr(b)$ under $\Inn(T)$.
Such orbit is therefore a conjugation invariant. 

Similar considerations apply to crossed modules. Let $(G,H)$ be a crossed module. 
Prop. \ref{prop:hitr2} suggests the following definition.

\begin{defi} \label{def:gentr2}
A trace $(\tr_b,tr_f)$ over $(G,H)$ is an augmented pointed quandle crossed module morphism of the 
conjugation quandle $\Conj(G,H)$ into some target augmented pointed quandle crossed module $(T,U)$. 
\end{defi}

\noindent 
When $(T,U)$ is the trivial pointed quandle $\Triv(\mathbb{C}_b,\mathbb{C}_f)$, we recover the restricted 
notion of trace of def. \ref{def:hitr1}.

By \eqref{qumor1}, \eqref{qucromor3}, $(\tr_b,\tr_f)$ being a trace over $(G,H)$ implies that
\begin{subequations}
\label{gentr2,3}
\begin{align}
&\tr_b(a\varrhd b)=\iota_{\tr_b(a)}(\tr_b(b)),
\vphantom{\Big]}
\label{gentr2}
\\
&\tr_f(a\varrhd B)=\mu_{\tr_b(a)}(\tr_f(B))
\vphantom{\Big]}
\label{gentr3}
\end{align}
\end{subequations}
with $a,b\in G$, $B\in H$, where $(\iota_x,\mu_x)$ denotes 
the inner automorphism of $(T,U)$ associated with $x\in T$ (cf. eq. \eqref{qucromor5,6}). Hence, 
for fixed $b\in G$, $B\in H$, $(\tr_b(a\varrhd b),\tr_f(a\varrhd B)$ with $a\in G$ varies in the 
orbit $\Orb_{\Inn}(\tr_b(b),\tr_f(B))$ of $(\tr_b(b),\tr_f(B))$ under 
$\Inn(T,U)$. Therefore, again, such orbit is a conjugation invariant. 

\vfil\eject

\end{document}